\documentclass[sigconf]{acmart}

\setcopyright{none}
\copyrightyear{2023}
\acmYear{2023}
\acmDOI{XXXXXXX.XXXXXXX}
\acmISBN{}
\acmConference[xxx]{Conference submission due}{2024}{}
\usepackage{etoolbox}
\patchcmd{\quote}{\rightmargin}{\leftmargin 2em \rightmargin}{}{}

\usepackage{todonotes}

\begin{document}
	
	
\title{What's going on at the back-end?\\
Risks and benefits of smart toilets}

\author{Isabel Wagner}
\email{isabel.wagner@unibas.ch}
\affiliation{%
    \institution{University of Basel}
    \city{Basel}
    \postcode{4051}
    \streetaddress{Spiegelgasse 5}
    \country{Switzerland}
}
\orcid{0000-0003-0242-6278}

\author{Eerke Boiten}
\email{eerke.boiten@dmu.ac.uk}
\orcid{}
\affiliation{%
	\institution{De Montfort University}
	\streetaddress{The Gateway}
	\city{Leicester}
	\country{UK}
	\postcode{LE1 9BH}
}

\begin{abstract}
	This paper presents a thematic analysis of an expert focus group considering smart toilets that record health data. The themes that arise indicate risks, many of which could be mitigated but currently are not, suggesting health benefits for the moment override other concerns only in specific application contexts.
\end{abstract}


\maketitle
	
\section{Introduction}
``Smart Toilets'' have been envisaged and produced for a few years now.
Bill Gates in 2018 joined Chinese and Japanese industries at the Reinvented Toilet Expo in Beijing, pushing for "disruptive sanitation technologies" \cite{gates2018reinvented}. His motivations
were economical and ecological, aiming for maximal water reuse -- which in particular requires the detection and elimination of pathogens in waste water. Smart toilets are currently widely in use in Japan \cite{ft2023china},
mostly as integral parts of ``smart homes" and with ``smart" relating to functionality and connectedness rather
than to intrusive functionality. These collect little data
on their users, other than biometrics to allow individualised customisation of services.

While there are risks associated with smart homes and IoT (Internet of Things) devices in general, this paper focuses on smart toilets that by their design collect significant data on their users. Smart toilets are typically developed from a medical perspective, broadly speaking as part of two future directions of medicine.

``\emph{Precision medicine}'' \cite{gambhir2021precision} aims for the best possible individualised treatment of patients by collecting maximal health and lifestyle data, normally in conjunction with AI-based processing. Measurement of toilet behaviour and analysis of excreta are known to provide meaningful health data \cite{ge2023passive}.

``\emph{E-Health}'' targets the remote monitoring of patients and delivery of healthcare, aiming to give patients more and longer independence, as well as seeking to cut costs of healthcare of an ever aging population. An overview of health-related IoT applications is included in \cite{mittelstadt2017ethics}.
The daily rhythm of toilet visits means that smart toilets provide an opportunity for regular health
measurements beyond those that are directly connected (e.g.\ excreta and urine flow), including heart rate and blood pressure \cite{medtech2019heart}. These would otherwise require wearable devices, which may reduce comfort, or separate actions which may be less reliably scheduled by the patient in their home. 

In addition to these two main medical drivers, smart toilets have also been considered for use in epidemiology, e.g.\ Covid testing \cite{ge2022covid}.

To reliably attribute recorded data to individuals, smart toilets need authentication, which is typically provided
via biometrics -- using either traditional or context-specific novel methods, such as anal print \cite{park2020mountable} or toilet seating habits \cite{zhang2021AItoilet}. For neither of these have we been able to find academic literature analysing their suitability as a biometric.

This paper investigates the risks and benefits of smart toilets that generate health data, in the spirit of a GDPR data protection impact assessment, looking at wide risks to rights and freedoms that follow from the data processing involved. The analysis
is based on an expert focus group that considered three specific smart toilet designs.

\section{Background and related work}
We focus on three toilet designs from the academic literature that have progressed to actual practical implementation. All of them send their recorded data to the cloud where it is then analysed.
Short descriptions of each, as shared with the focus group participants, are included in Appendix \ref{app:scenarios}. 

The Heart Seat\mbox{${}^{\mathrm{TM}}$} \cite{conn2019in-home} records heart rate, blood pressure, and other measurements related to heart health.
The smart toilet developed by startup company Coprata based on research
at Duke University ({\tt https://smarttoilet.pratt.duke.edu/research}) \cite{zhou2021stool} takes stool images after flushing.
The ``Precision Health Toilet" developed by a group based at Stanford University \cite{park2020mountable} has four cameras, for authentication and recording of a wide range of toilet behaviour and excreta.

A recent analysis by the Stanford group (Ge et al.\ \cite{ge2023passive}) touches on many risks of smart toilets. These include data protection and privacy concerns, particularly for children and in the light of cultural diversity, as well as secondary use risks of collected data.

While the research literature on specific risks of smart toilets is very limited, many of the known
privacy and security risks of IoT products obviously also apply in this specific context.
A survey of ethical issues around health IoT is given by Mittelstadt \cite{mittelstadt2017ethics}. In particular how health IoT devices may principally impact physical and informational privacy is addressed in great detail in this paper.

Most smart toilets, including the three examples discussed here, record their data in the cloud. There have been serious data leaks from cloud services, through malice, negligence, or combinations of both. High-profile examples of this include the Ashley Madison, Sony, Yahoo, and Equifax breaches \cite{saleem2020sok}.

Data subjects in IoT-collected data may also be at risk from those who collect or have legitimate access to the data. Data collected may be excessive, giving intrusive views of people's habits, that can be extrapolated
to more general behaviour or monetised. Many IoT devices already track users' behaviour well beyond their users' interests, e.g. smart TVs
\cite{moghaddam2019watching,varmarken2020tv}.
IoT-collected data may also be sent to third parties, including sometimes in plaintext \cite{ren2019information}.
Third parties observing internet traffic may draw inferences of user behavior and device ownership from this.
Ownership may be stigmatising, or imply a security and safety risk in the case of medical devices.

Similarly, enumerating IoT devices over the Internet enables for vulnerable ones to be identified, and also to infer who owns which IoT devices \cite{perdisci2020iotfinder}.
Insecure IoT devices may be used to spy on their owners, e.g., with smart speakers or baby monitoring cameras, or to use them in a botnet, e.g., Mirai \cite{xenofontos2022consumer}.

While according to \cite{baranchuk2018cybersecurity} there is a lack of evidence that hacking is a relevant clinical problem, \emph{security} issues of existing medical products have consequences extending beyond confidentiality-related problems. Attacks on availability or data integrity on networked pacemakers \cite{marin2016insecurity,baranchuk2018cybersecurity}, insulin pumps \cite{li2011hijacking}, etc., give way to \emph{safety} problems, i.e.\ they may put human lives at risk.

\section{Methodology}
To further the analysis of the impact of smart toilets beyond what is already represented in the academic literature as described above, we used a \emph{focus group} of experts. They were recruited from a community of specialists grown from Twitter interactions initially through a series of face-to-face round table events (advertised open-to-all on social media) which had been contributing significantly to our previous research, mostly on privacy impact assessment \cite{ferra2020challenges}. While for this research only three experts participated, 
 (see Table \ref{tab:participants}), this represents a huge amount of expertise and experience. Each of them
has at least a decade of professional experience in privacy and data protection, including advising on and leading impact assessments; in addition, between them they add significant experience in information security, law, and medicine and medical practice.

\begin{table}
	\caption{Participants and their expertise}
	\label{tab:participants}
	\begin{tabular}{lp{6cm}}
		Participant & Expertise \\
        \midrule
		P1 & Data protection officer for local government authority \\
		P2 & Privacy and data protection lawyer with medical degree and experience\\
		P3 & Privacy and data protection consultant with security expertise \\
	\end{tabular}
\end{table}

In addition to participant information and consent sheets,
the participants were given scenarios of practical designs related to the academic literature (included in full in Appendix \ref{app:scenarios}) and a list of questions drawn up by the researchers (see Appendix \ref{app:questions}). The questions are in the spirit of GDPR data protection impact assessment, focusing on ``what could possibly go wrong", aiming to identify wide-ranging risks to rights and freedoms that follow from the data processing involved, i.e.\ well beyond a compliance-oriented view of data protection.

Discussions were recorded and then transcribed. Subsequently
we applied thematic analysis \cite{braun2006using} to extract themes.

Our analysis has resulted in three overarching themes, which are described in the subsequent section:
\textit{functionality and perception} is about what smart toilets do, how they are designed, and how they may be perceived by the public (Section \ref{sec:functionality});
\textit{data protection and security} is about data practices including choice and control for users, as well as security issues including authentication and data confidentiality (Section \ref{sec:privacy});
and \textit{effects and side-effects} concerns the benefits of smart toilets as well as negative effects they may cause, including medical effects, discriminatory effects, mandated or coerced use, and effects on specific population groups (Section \ref{sec:effects}).
Each of these sections describes the themes elicited from the focus group discussion, illustrated with representative quotes from the participants.

\section{Smart toilets: functionality and perception}
\label{sec:functionality}
This overarching theme gathers issues directly related to the operation of these smart toilets and the public's
perception of them.
\subsection{Features}
A smart toilet designed to discover health issues,
maybe even to the extent of it being a medical device in a legal sense, may lead to an issue of liability: who is responsible for making sense of the recorded data, in particular, is the manufacturer liable if the recorded data miss a real health issue?
\begin{quote}
	``And if the toilet doesn't detect something, are the manufacturers liable?'' (P3)
\end{quote}

\begin{quote}
	``Presumably, all they're doing is recording the data and they're outsourcing the headache that comes with any abnormalities to you and your own healthcare provider.'' (P2)
\end{quote}

There may be operating costs beyond the initial purchase, -- will the toilets need regular maintenance, e.g., in terms of refilling reagents? How much power will they need?

\begin{quote}
	``it’s got a urinalysis,[...] there's going to be maintenance and upkeep'' (P1)
\end{quote}

\begin{quote}
	``Cost of energy crisis.'' (P3)
\end{quote}

\subsection{Design process}
For innovative devices, there is normally a question of what has driven the design: an actual perceived need (in this case, a health problem), or an economic opportunity?

\begin{quote}
	``It's that typical, ‘here’s some shiny tech. Let's throw it at a problem that we don't even know if we need to solve in this way.’'' (P1)
\end{quote}

\begin{quote}
	``there's an incentive to monetize everything instead of actually solving the problems, because those two are not necessarily in alignment'' (P3)
\end{quote}

Particularly for devices that solve a problem using potentially invasive data processing, like here,
data protection law asks questions of necessity and proportionality. To what extent have alternative solutions been explored (and found inferior)? 
Is there really a need for collecting and processing this data?

\begin{quote}
	``they don't look like horrible ideas unless you know how the world of data and technology and power dynamics works'' (P3)
\end{quote}

\begin{quote}
	``I would want to see evidence that there was a whole range of potential ideas looked at, and why this is the best; and the answer, ‘because it will make us the most money,’ is not eligible'' (P3)
\end{quote}

Solutions based on data collection should have privacy preserving approaches built in -- for moral reasons, because they exist; for legal reasons (e.g.\ \textit{privacy by design}  under GDPR), and to reduce security risks. It is important to take into consideration that the operator/manufacturer of the product should be considered a potential adversary, who might e.g.\ consider re-sharing of data, or function creep using data they do not really need to obtain in the first place. Risks do not just arise from third parties attacking the system.

\begin{quote}
	``pseudonymize that data that they’re storing'' (P1)
\end{quote}

\begin{quote}
	``You could drive the privacy by design concept by default into the production, which is clearly not part of the package in these instances.'' (P2)
\end{quote}

\begin{quote}
	``it always comes back to the same thing in data protection and privacy, doesn't it? It's all very well when the good guys are doing it, but you have to think about what would happen if this was being operated by the bad guys.'' (P1)
\end{quote}

The need for privacy by design puts an onus on product designers to have awareness of the available methods.

\begin{quote}
	``cack-handedness in terms of design by people who might be very technically competent in sensors, but don't really understand the implications of the data risks.'' (P2)
\end{quote}

\begin{quote}
	``the basic premise that we've got all of these sensors now. They're small. They're easy to manufacture. There are lots of people out there just thinking, ‘well, what can we stick them in? And badge in some way that makes it work -- `it's going to be for your health'.'' (P2)
\end{quote}

Product designers may have very good intentions, but this actually complicates the problem of getting privacy protections into products.

\begin{quote}
	``people with the noblest of intentions are probably the least likely to critically self-analyse.'' (P3)
\end{quote}

\begin{quote}
	``if you try and point out the risks, then they get terribly offended and defensive'' (P3)
\end{quote}

\subsection{Data versus knowledge}
The three smart toilets in our study started from medical research where the medics aimed to measure relevant diagnostic data. This is not obvious from their descriptions,
and not true in general for health oriented IoT devices including smart toilets.
They may well record data in large quantities, but this data need not correspond to a medical diagnosis, or even necessarily information about medical conditions.

\begin{quote}
	``what you do through this sort of technology is you identify anomalies. You have no real sense of what those anomalies mean, or what their implications are in real health terms'' (P2)
\end{quote}

\begin{quote}
	``they report back to you your bowel habit is abnormal because it lies outside of the 95\% normal range. Now over to you. Do whatever you want with that.'' (P2)
\end{quote}

Sensing and recording large amounts of data is easy, cheap, and easy to automate. However, interpretation and making sense of the data requires much more significant investment.

\begin{quote}
	``the algorithm for analysis, is there any human in the loop or are they just feeding everybody's data and expecting it to come up with conclusions that may or may not reflect?'' (P3)
\end{quote}

\begin{quote}
	``is it a person looking at that?'' (P2) -- ``And are they qualified?'' (P3)
\end{quote}

\begin{quote}
	``will they be farming that out to call centres with scripts and checklists?'' (P3)
\end{quote}

In addition to individuals getting information about their health, the recorded data can allow far more wide-ranging inferences, not just about individuals but also their families.

\begin{quote}
	``presumably you could infer data in relation to people's diets [...] their routine as far as that is spent in their home.'' (P2)
\end{quote}

\begin{quote}
	``They might have enough data that they've pieced together about me to know that heart disease runs in my family or dementia runs in my family'' (P1)
\end{quote}

\subsection{Public perception and understanding}

Transparency on what happens with the recorded data is likely out of reach of the users. Experts worry about data practices
in a way a typical user would not -- even if some information on this is available, most users likely will not access it.

\begin{quote}
	``Most people who look at that technology think, ‘well, I'm wearing the watch. I'm seeing the data. That's all that's happening,’ and are completely blind to even the possibility that it's being hoovered up and transmitted elsewhere and used for other purposes'' (P2)
\end{quote}

\begin{quote}
	``your first thought would not be for most people, ‘what's happening to the data?’'' (P2)
\end{quote}

\begin{quote}
	``Joe Bloggs on the street who maybe doesn’t understand some of it and doesn't read privacy notices'' (P1)
\end{quote}

Smart toilets are likely to hit a wide range of 
expectations of privacy and perceptions of data sensitivity.

\begin{quote}
	``lots of people will be happy with Alexa and Ring doorbells, but yeah, taking pictures of your bum on the toilet, I think most people go, ‘what? That's too far.’'' (P1)
\end{quote}

\begin{quote}
	``Some people would be horrified by that. Some people will be like, ‘I want one. I need to check whether I'm ill or not,’'' (P1)
\end{quote}

\begin{quote}
	``who hasn't photocopied their bottom at an office party? But this is taking it a bit far.'' (P3)
\end{quote}

However, clever marketing or a perceived need can override privacy concerns, even in privacy professionals.

\begin{quote}
	``They [people who work in data protection] want the app, they download it. I think this is similar, isn't it?'' (P1)
\end{quote}

\begin{quote}
	``we've seen that with all smart technology, whether it's your Ring doorbell, or your Echo or your Alexa, people are like, ‘ooh, shiny gadget. I want one of them. That's really good.’'' (P1)
\end{quote}

Potential users have no intrinsic reason to trust the manufacturers of smart toilets. However, if smart toilets are introduced by a trusted provider, assurances about data sharing may be received better and privacy concerns can be alleviated.

\begin{quote}
	``would people feel differently if this was your doctor saying, ‘we can issue you with one of these toilets on the NHS\footnote{NHS (National Health Service) is the publicly funded healthcare system in the UK.} and the data just comes back to the NHS,’ as opposed to, ‘you could buy one of these and the data is processed in this cloud by this private company,’?'' (P1)
\end{quote}

\begin{quote}
	``If you could have confidence that the data was simply being used for the purpose that you would naturally intuit it was going to be used for, telling you what was going on or providing that information to your trusted healthcare professional, people would be relatively content with that'' (P2)
\end{quote}

However, people may have limited trust in private corporations, especially when it concerns handling of health data.

\begin{quote}
	``lack of trust in authority and a definite lack of trust in private health corporations, especially Americans, which is all about making money out of your health data'' (P1)
\end{quote}

On the other hand, people are often quite trusting in health care situations, assuming devices are safe because they are allowed to be sold, or prescribed.

\begin{quote}
	``I assume that just because it's being offered for sale, everything about it must therefore be safe and lawful, for which I have two words -- \textit{radium condoms}.'' (P3)
\end{quote}

Altogether, potential users need to balance benefits
against risks. They may well be aware of economic costs and privacy impacts, but these may be worthwhile trade-offs against a health benefit. This is likely to be an individualistic choice, but especially individuals with existing health conditions and older individuals may lean towards the health benefit. This also indicates that smart toilet providers have (or should have) a particular duty of care to protect user privacy.

\begin{quote}
	``[if] you're already a higher risk of harm, that might make you give up some of those privacy risks if it means it’ll save your life one day.'' (P1)
\end{quote}

\begin{quote}
	``you probably need a bit of a risk assessment of whether it suits the individual, don’t you? If you're going to have these privacy risks, do they outweigh the health risks?'' (P1)
\end{quote}

\begin{quote}
	``Is this all going to be worth it for me as an individual to pay a lot of money and give up lots of potential privacy?'' (P2)
\end{quote}

In line with literature reports of \textit{alert fatigue} from health IoT devices, feedback from 
toilets sold as lifestyle or wellness products might be repetitive, and thereby become an annoyance.

\begin{quote}
	``I think I’d just get annoyed by it. [...] If my toilet every single day is going, ‘drink more water, eat more fruit and veg, cut down on the processed food and the sugars,’ I'll just switch it off. I know all those things; my toilet doesn't need to tell me.'' (P1)
\end{quote}

Smart toilets also impact on social interactions.
Any kind of smart technology in the home raises questions of transparency and consent when guests come into the home. For most people, smart toilets may be more awkward to explain to their guests than, for example, smart speakers

\begin{quote}
	``Can you imagine your friends coming around for dinner? And they look at your toilet and they're like, ‘what's that?’ ‘They're just cameras. They’re just looking at your bum.’'' (P1)
\end{quote}

Feedback from smart toilets may also influence social interactions with healthcare providers and doctors.

\begin{quote}
	``And will I start crying wolf and will the GP keep going, ‘God, she’s back again because the toilet’s told her again. There’s nothing wrong with her?’'' (P1)
\end{quote}

\begin{quote}
	``the doctor perhaps has the same attitude towards patients researching their own symptoms and instantly goes, ‘this is a mental health problem, not a physical one.’'' (P3)
\end{quote}

\section{Data protection, privacy, and security}
\label{sec:privacy}
Several themes arose from the focus groups in relation to data protection, privacy and security, in addition to necessity and perceived privacy impacts (already considered above).

\subsection{Choice and control}
One important dimension of privacy is an individual's control
of what is shared about them and with whom. 

Smart toilets could and should offer such choices, for example at configuration during setup. However, rather than giving users genuine control, this may also lead to cognitive overload.

\begin{quote}
	``Do you want to use it offline? Do you want it to upload automatically? Do you want to send the data to A, B, C, D, or a combination?'' (P1)
\end{quote}

\begin{quote}
	``That sort of thing is a massive cognitive load, and after about five different kind of questions and settings, people are just going to be like, `urgh.''' (P3)
\end{quote}

An obvious choice that may or may not be available is whether the \textit{smart} functionality can be avoided, using just the functionality of the \textit{toilet} by itself.

\begin{quote}
	``Can you flush it if you haven't done fingerprint recognition? How much choice do you get in real terms?'' (P2)
\end{quote}
The choice does not just apply to being able to flush without authenticating first, it is also about allowing use of the toilet without triggering data collection (which may be done by sensors hidden in the plumbing).

Medical data should not be stored centrally (e.g.\ on the cloud) in an identifiable form; decentralised storage under the user's control gives alternative options.
In particular, privacy preservation and data minimisation can benefit from local processing of the data, for example on the user's phone.

\begin{quote}
	``It's not collating that data in an identifiable form centrally, it's about giving you the data to do with it what you want to do as an individual, and then it's up to you who you share it with.'' (P1)
\end{quote}

\begin{quote}
	``I don't want people to have photographs of my poo in the cloud'' (P1)
\end{quote}

\begin{quote}
	``why not just pair it with your phone instead of having to send in [data]'' (P3)
\end{quote}

The acceptability of cloud storage depends on who the cloud entity is, e.g., a private company vs.\ a public health provider.

\begin{quote}
	``depending on where the data's going. Is it going to the cloud? Is it a private company? Is it under contract with the NHS?'' (P1)
\end{quote}

\subsection{Authentication}

Authentication is relevant because toilets may be used by more than one person, including guests who do not know that this is a smart toilet and have not consented to any data collection.

\begin{quote}
	``How does the toilet know it's you?'' (P3)
\end{quote}

\begin{quote}
	``what happens if other people sit on it?'' (P1)
\end{quote}

Authentication options for toilets in the study were fingerprint and anal print.
For anal print \cite{park2020mountable}, it is not clear that it can function as a responsible biometric.

\begin{quote}
	``anal print is used for authentication, and you suddenly come down with a bad case of piles. Does that mean you can't [use the toilet]?'' (P3)
\end{quote}

It is also unclear what is involved in the enrollment of the anal
print.

\begin{quote}
	``when you have to do your thumbprint on your phone, when you set it up and you have to do a few different ones, do you think you have to like to sit around on the toilet seat a few times like that to get it right?'' (P1)
\end{quote}

\subsection{Access Control}

Especially in multiple occupancy homes or settings like nursing homes, the device may be installed by someone who
has no legal authority over the users' personal data. In that situation, it is worth asking who controls the accounts, and who has access to the collected data.
Would the purchaser of the device have access to all users' data?

\begin{quote}
	``who holds the account that has access to all the data? Who sets up a new user?'' (P2)
\end{quote}

\subsection{Confidentiality and integrity}
Local data storage, cloud storage, and transfer to the cloud
all incur risks of data loss, both of identifiable medical data and user account data.
Data loss is a common occurrence in general.

\begin{quote}
	``the data being released onto the public domain'' (P1)
\end{quote}

\begin{quote}
	``your password is exposed in association with your email address, that's a risk that's got nothing to do with your toilet'' (P2)
\end{quote}

There may also be a risk of hacking into the toilet. The toilet may be insufficiently secure
-- in general, insecure IoT products are very common, and integrity is a
real issue for health devices.

\begin{quote}
	``the amount of fear you could cause somebody from hacking their smart toilet and having it tell them that they were dying, or even having it tell them that they were fine when they are actually ill.'' (P3)
\end{quote}

\subsection{Data practices}
The data generated by smart toilets is sufficiently rich and extensive that it raises concerns about where it goes, and whether there is purpose limitation in place.

\begin{quote}
	``who is going to get the data and what else are they going to do with it?'' (P1)
\end{quote}

\begin{quote}
	``private companies and insurance companies in particular, getting their hands on our health data'' (P1)
\end{quote}

Data practices can also change over time, and it is hard for users to gain assurance that the promises regarding data practices made when the smart toilet is purchased are being kept.

\begin{quote}
	``how do they ever get to a point where they could confidently walk into a shop and buy one and know that there is none of this murky stuff designed into it?'' (P2)
\end{quote}

\begin{quote}
	``some sort of certification scheme: [...] We have audited and reviewed their privacy practices and we are satisfied in fact that the data is only used for the purposes that you would envisage.'' (P2)
\end{quote}

Data sharing could be incentivized: if data processing and sharing is based on user consent, then smart toilet makers as well as other entities (e.g., insurance companies) could design incentives to make more users consent.

\begin{quote}
	``not `we will increase your premium,' but, `if you install one, we will decrease your premium.''' (P2)
\end{quote}

\begin{quote}
	``it becomes manufactured consent, because you don’t always have the choice to say no'' (P3)
\end{quote}

The smart toilet makers' business models may depend on monetizing the health data generated directly, or by using it for follow-up sales of remedies.

\begin{quote}
	``You have no real sense of what those anomalies mean, or what their implications are in real health terms, but they afford an opportunity to sell stuff'' (P2)
\end{quote}

\begin{quote}
	``smart technology is just the new radium. Everyone's rushing to monetize it and commercialize it, and the people buying it are assuming a good faith on the part of the sellers that simply isn't there.'' (P3)
\end{quote}

\section{Effects and side-effects}
\label{sec:effects}
\subsection{Benefits}

Alongside what could possibly go wrong, potential benefits of smart toilets also emerged from the focus group discussion.

Data can be used to help people and prolong or improve their lives.

\begin{quote}
	``you can now use so much data to predict, to prevent, to help people'' (P1)
\end{quote}

\begin{quote}
	``for some people, this would probably be really good because they don't have to think about it every day. They do go to the toilet automatically every day, and it does it for them, particularly people with memory loss and dementia and elderly people, etcetera'' (P1)
\end{quote}

\begin{quote}
	``If I've already got a heart problem, I'm going to do everything I can to monitor myself and take care of myself'' (P1)
\end{quote}

Some people enjoy having data about themselves to look at.

\begin{quote}
	``there is a cohort of people out there who are obsessed about self-quantification and having information'' (P2)
\end{quote}

\subsection{Medical effects}

The development of these smart toilets is driven by the desire to find a health benefit. However, the specific
context raises many questions on how this is achieved and possible side effects of this approach.

First, it needs to be clear that there is indeed a demonstrable health benefit. Is there evidence that the recorded data can support medical conclusions or decisions?
Far more data is recorded on people with no acute serious medical problem than would be in traditional health care.

\begin{quote}
	``what you do through this sort of technology is you identify anomalies. You have no real sense of what those anomalies mean, or what their implications are in real health terms'' (P2)
\end{quote}

\begin{quote}
	``does it actually make a difference to know what your blood pressure is every minute of every day?'' (P2)
\end{quote}

\begin{quote}
	``you've now had a brief episode of what looks like a-fib [atrial fibrillation] once, because we've been monitoring your heart for 24 hours a day for the last three months. Is that of any significance?'' (P2)
\end{quote}

People may remain acutely aware that they are constantly being monitored, and this could induce 
anxiety.

\begin{quote}
	``is all I'm going to be doing making myself anxious because I'm going to be seeing once every three weeks that I've got something funny, a blip, or my blood pressure has gone down and now I'm panicked about that?'' (P2)
\end{quote}

\begin{quote}
	``You’d just think you were dying in every day.'' (P1)
\end{quote}

\begin{quote}
	``the real risk here is about that over-surveillance, anxiety-inducing, problem-finding, what medics call incidental illness.'' (P2)
\end{quote}

\begin{quote}
	``you put clinicians in a very difficult position because they'll say, ‘well, I've got an anxious patient now. Their anxiety needs to be addressed. There's some vague indication of something abnormal and the only routes that I now have to take this any further are invasive tests, which carry risk.'' (P2)
\end{quote}

\begin{quote}
	``you could see for some people this sort of feedback being incredibly stressful because I'm not pleasing my toilet. My toilet is a disappointed in me.'' (P2)
\end{quote}

People may end up framing their health in terms of what sensors can measure, possibly with an undue focus on one part of their health system.

\begin{quote}
	``if it's kind of framing metrics in terms of heart health, you could very well miss a pituitary tumour that causes the same effects, but because there's a bias towards framing it in terms of one system, that could mean failing to, or misleading a diagnosis of another system'' (P3)
\end{quote}

\begin{quote}
	``Or it says you need to be eating more vitamin D, and you eat a hell of a lot of vitamin D, and it still says you need to be eating more vitamin D and you go, ‘well, it must be faulty,’ and don't consider the fact that you might have a tapeworm'' (P3)
\end{quote}


The quantification of health may also interact with the framing of health and wellness as something that individuals have a personal responsibility and choice in -- especially in areas where that does not fairly reflect their reality. 

\begin{quote}
	``It also generates this idea that health and wellness has to do with metrics, and you’re well if your metrics tell you you're well.'' (P2)
\end{quote}

\begin{quote}
	``I have a chronic genetic condition. I'm always a bit broken. Am I healthy? And can I improve my health? Well, no, except by changing my genes'' (P3)
\end{quote}

\begin{quote}
	``There are so many things about health and wellbeing that individuals don't have any control over. There’s genetics. There's the environmental conditions. There's the life circumstances.'' (P3)
\end{quote}

The increasing amount of diagnostic information produced by health devices may lead to an increasing load on the health system.

\begin{quote}
	``I’d be at the doctors every few days, I think.'' (P1)
\end{quote}

\begin{quote}
	``what good is detecting signs of illness when the infrastructure to treat them is already overloaded?'' (P3)
\end{quote}

\begin{quote}
	``Eventually, the only thing you can do is a potentially dangerous investigation like a colonoscopy because your toilet has told you. [...] So, are you going to see a spike in colonoscopies because that's the only solution for this problem? It’s created a massive increase in the input into the system. As a consequence, you’re getting a massive increase in the number of false negatives, of negative colonoscopies, but you're exposing the patients to risk, and eventually one of them is going to have a perforation or suffer a serious adverse consequence'' (P2)
\end{quote}

\begin{quote}
	``they're outsourcing the headache that comes with any abnormalities to you and your own healthcare provider.'' (P2)
\end{quote}

\begin{quote}
	``if it is genuinely saying to me, ‘hey, you might have cancer,’ great. I need to go as early as possible. Brilliant, fantastic news. But how accurate is that prognosis? And then am I just wasting everybody's time.'' (P1)
\end{quote}


Where devices observe ``anomalies'', there are risks associated with where the baselines for such considerations come from, and how accurate or biased they are.

\begin{quote}
	``if you buy one of these things and you've already got cancer and you don't know it, it's going to build a baseline on a state which you don't want to be in in the first place.'' (P3)
\end{quote}

\begin{quote}
	``there's quite a wide range of variability in how human bodies function and at the moment, anyway, the baselines are all taken from white males.'' (P3)
\end{quote}

\begin{quote}
	``Have they been tested to make sure they work for dark skin, because most skin sensor technology doesn't?'' (P3)
\end{quote}

There may be little or even a negative health benefit, due to bias or undue focus on the smart toilet data analysis.

\begin{quote}
	``The health yield is going to be pretty low if there is any at all, because it’s going to be people that have the money who are seduced by technology, who are probably comparatively young and healthy'' (P2)
\end{quote}

\begin{quote}
	``Or somebody's feeling dodgy, but their loo says they’re fine, so they don't go to the doctor'' (P3)
\end{quote}

\begin{quote}
	``The false reassurance, because your toilet’s telling you that you need to eat more fibre and you think, ‘well, that's the solution. More fibre,’ but in fact, what you really need to be doing is taking whatever symptoms you have seriously and getting it checked out.'' (P2)
\end{quote}

\subsection{Discriminatory effects}

When extra health analysis and more diagnosis is introduced, this could lead to
discrimination against people with illnesses, e.g., through higher insurance premiums.

\begin{quote}
	``I’m not aware that when I get that quote, they've bumped it up because they actually know from my data I’m at risk'' (P1)
\end{quote}

\begin{quote}
	``those that have more ailments are going to be at a disadvantage economically.'' (P1)
\end{quote}

\begin{quote}
	``Or you run a nursing home and you're doing a cost-benefit analysis on the types of residents you want to take in, so you run a study from all this clever, smart data that you have the account details of because you were the purchaser for all these things, and you decide that people with these particular conditions are a better return on investment.'' (P3)
\end{quote}

\begin{quote}
	``It becomes part of the standard interviewing process. [...] 'would you just like to use the toilet on your way out?''' (P2)
\end{quote}

Discrimination against people with dark skin tones is also a worry.
\begin{quote}
	``Have they been tested to make sure they work for dark skin, because most skin sensor technology doesn't?'' (P3)
\end{quote}

\subsection{Mandated or coerced use}

The use of smart toilets should be with full consent and knowledge.
Covert or mandated use  by employers or other organizations could lead to adverse scenarios.

\begin{quote}
	``it's entirely possible that that could be installed without the user knowing, and then when your employer or your insurance company or your landlord or your bank, or whoever, says that these are necessary security vetting checks, and you can't have your whatever without it, and the creeping incremental intrusion and loss of autonomy is pretty major.'' (P3)
\end{quote}

\begin{quote}
	``if a care home installs it, for example, and they don't have any choice; that's the toilet that they've got to use on their floor in the care home?'' (P1)
\end{quote}

\begin{quote}
	``with the fingerprint recognition, if you install that in the offices, then you have a way of monitoring whether somebody who says they're on a toilet break is actually on a toilet break.'' (P2)
\end{quote}

There is precedent for this, not yet at the individualized level of smart toilets, but indicating that there may be substantial institutional demand for smart toilets.

\begin{quote}
	``There are university campuses in America where they have detectors in the sewage system to sound the alarm if certain metabolites or certain drugs are picked up'' (P3)
\end{quote}

\begin{quote}
	``they’ll absolutely be advertising these to like the military with that specific use case in mind, probably to education as well.'' (P3)
\end{quote}

\begin{quote}
	``I think the uptake will be in the areas where the individuals don't get to make the choice for themselves.'' (P3)
\end{quote}

\begin{quote}
	``I can actually see this not being terribly popular in the consumer market but being extremely popular in the institutional market and that's bad.'' (P3)
\end{quote}

\subsection{Effects on specific groups}

Elderly people and people with specific conditions might benefit more.

\begin{quote}
	``If I’m 87 with a lot of illnesses, maybe I'm past caring about anybody knowing my anal print at that point if it means I get another few years.'' (P1)
\end{quote}

\begin{quote}
	``for some people, this would probably be really good because they don't have to think about it every day [...], particularly people with memory loss and dementia and elderly people, etcetera''
\end{quote}

Caution is required to preserve the
rights of people who lack capacity.

\begin{quote}
	``people that are lacking capacity, where you've got elderly people, vulnerable people and children, etcetera. What rights have they got with this sort of technology if a care home installs it, for example, and they don't have any choice; that's the toilet that they've got to use on their floor in the care home?'' (P1)
\end{quote}

\section{Conclusion}
\begin{quote}
``From a societal point of view, if it helps a tiny minority of people and harms the majority of people, is that a case for saying, ‘do you know what? Actually, we're not going to encourage this sort of thing. We shouldn't be doing it.’'' (P3)
\end{quote}

It is clear that smart toilets can be extremely beneficial in some scenarios. This includes where there is a medical need for continuous monitoring of specific health parameters, or where other means of monitoring would be too onerous (e.g., time-consuming to travel to hospital) or error-prone for the patient (forgetting to take measurements).

The trade-off between privacy risk and broader risks,
and potential health benefit must be carefully considered.
It looks like for broad and general use, the health benefit is likely too low in view of privacy risks and negative effects they may cause.

However, smart toilet makers can take steps to address the privacy risks. They can implement technical measures, such as privacy enhancing techniques including local data processing and pseudonymization of recorded data. Two main issues with this are:
\begin{itemize}
    \item published designs and academic literature do not give the impression that there is sufficient expertise in privacy(-by-design) and data protection at the point where the toilets are designed;
    \item the business model most likely relies on exploitation of health data in some form, and therefore implementing privacy-preserving measures is not attractive for smart toilet makers.
\end{itemize}

Fitting smart toilets in with existing regulation on medical devices would address several of the concerns. 
Smart toilets should therefore be classed as medical devices and not sold as consumer products, due to numerous negative effects including systemic effects.

\bibliographystyle{ACM-Reference-Format}
\bibliography{smart-toilets}

\appendix

\section{Focus group questions}
\label{app:questions}

\begin{itemize}
	\item What could go wrong if this technology was widely adopted?
	\item What risks do you see with this technology?
	\item What would persuade you to use this technology?
	\item What would persuade you to recommend this technology in a professional context?
	\item What would persuade you to recommend this technology to your friends and family? (e.g., benefits, personal circumstances, privacy protections, or modifications to the product)
\end{itemize}

\section{Scenarios}
\label{app:scenarios}
See next page.

\begin{figure*}
	\includegraphics[width=.9\textwidth]{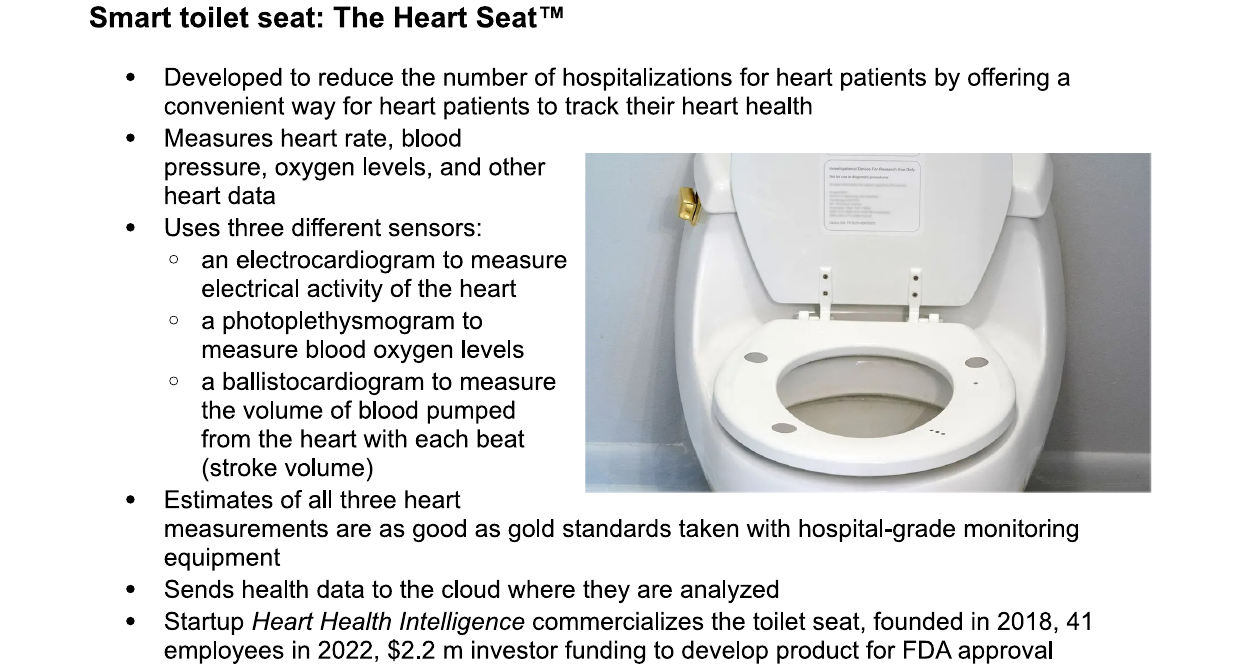}
	\caption{Scenario 1: smart toilet seat \textit{The Heart Seat} \cite{conn2019in-home}}
\end{figure*}

\begin{figure*}
	\includegraphics[width=.9\textwidth]{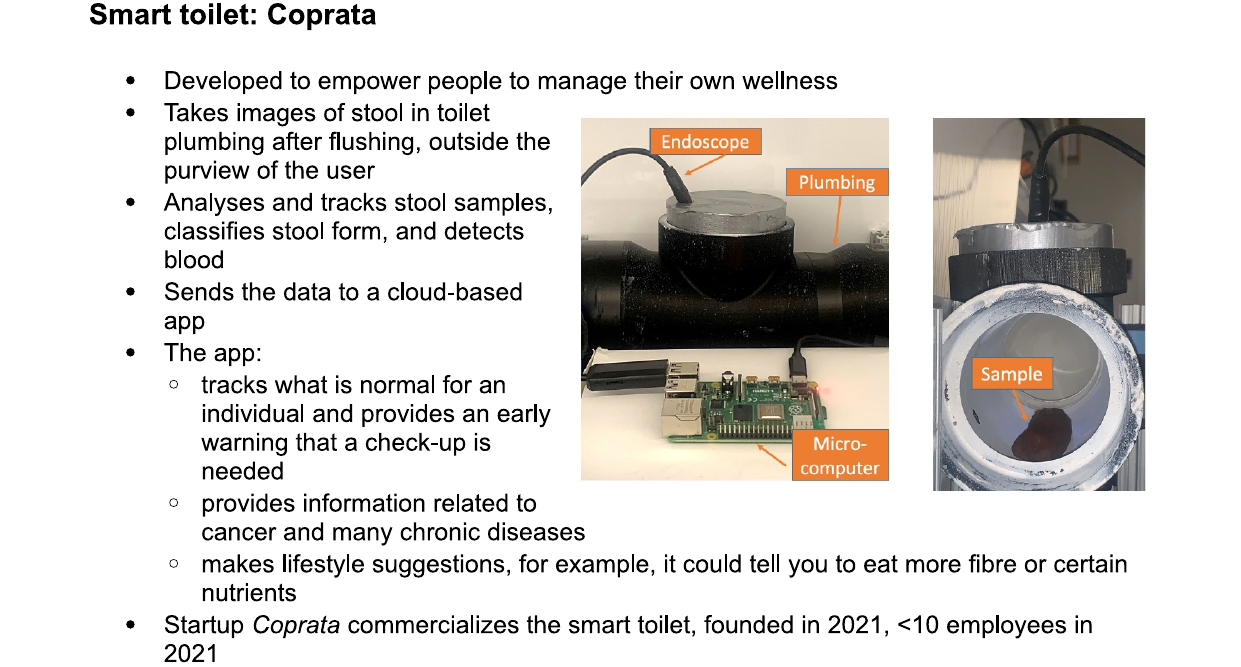}
	\caption{Scenario 2: smart toilet \textit{Coprata} \cite{zhou2021stool}}
\end{figure*}

\begin{figure*}
	\includegraphics[width=.9\textwidth]{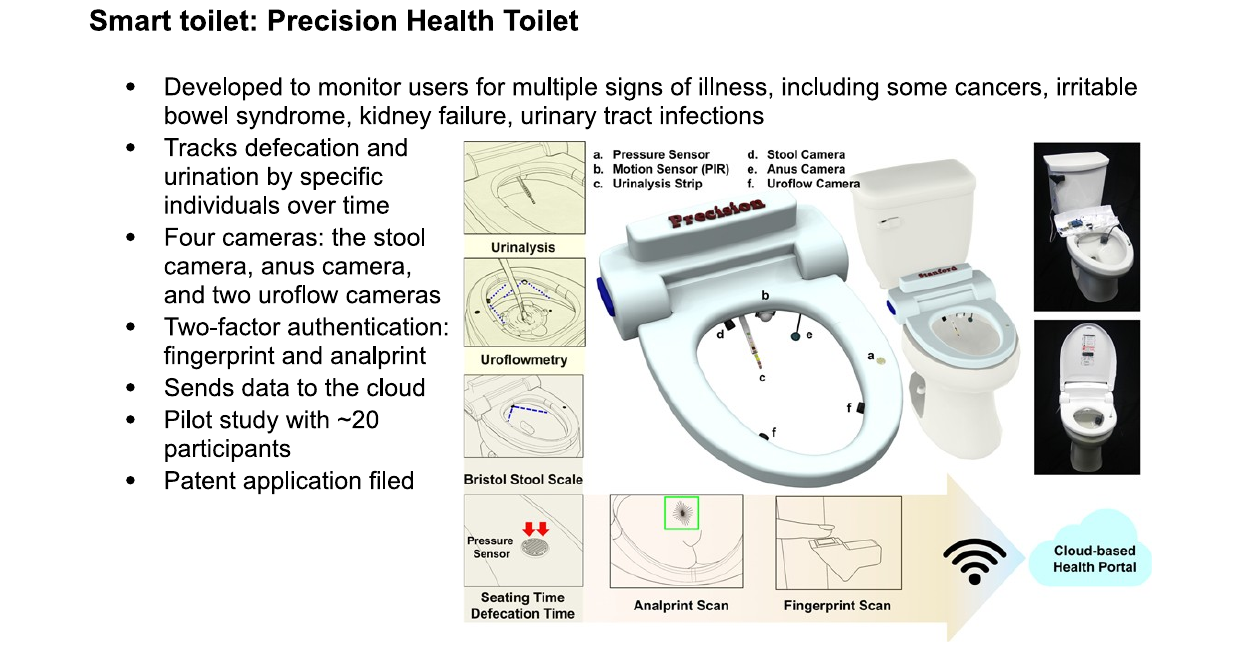}
	\caption{Scenario 3: smart toilet \textit{Precision Health Toilet} \cite{park2020mountable}}
\end{figure*}
\end{document}